\documentclass[9pt,twocolumn,twoside]{pnas-new}

\templatetype{pnasresearcharticle} 

\usepackage{gensymb}

\makeatletter
\fancypagestyle{firststyle}{
   \fancyfoot[R]{\footerfont PNAS\hspace{7pt}|\hspace{7pt}\textbf{\today}\hspace{7pt}|\hspace{7pt}vol. XXX\hspace{7pt}|\hspace{7pt}no. XX\hspace{7pt}|\hspace{7pt}\textbf{\thepage\textendash\pageref{LastPage}}}
   \fancyfoot[L]{\footerfont\@ifundefined{@doi}{}{\@doi}}
   \fancyhead{}
}

\makeatother

\begin{document}

\title{Unconventional Magneto-Optical Effects in Altermagnets}

\author[a,1]{Yongpan Li}
\author[a,1]{Yichen Liu}
\author[a,2]{Cheng-Cheng Liu}

\affil[a]{Centre for Quantum Physics, Key Laboratory of Advanced Optoelectronic Quantum Architecture and Measurement (MOE), School of Physics, Beijing Institute of Technology, Beijing 100081, China}

\leadauthor{Li}

\significancestatement{The conventional view is that magneto-optical effects (MOEs) are driven exclusively by Berry curvature, implying that ideal altermagnets with effective time-reversal symmetry should be magneto-optically inactive. Here, we challenge this understanding by identifying a new mechanism for MOEs: quantum metric-induced or -dominated responses, which we term unconventional MOEs. Furthermore, we derive general formulas that incorporate both Berry curvature and quantum metric for unconventional MOEs in altermagnets, enabling a quantitative evaluation of their respective contributions. Using symmetry analysis and first-principles calculations, we demonstrate that such unconventional MOEs occur in both ideal and realistic altermagnets. Our work not only establishes a correct and unified theory of MOEs in altermagnets, but also opens a new direction for exploring quantum geometry driven magneto-optical phenomena.}

\authorcontributions{Author contributions: Cheng-Cheng Liu initiated the idea. Yongpan Li and Yichen Liu carried out DFT calculations and formula deduction. All authors contributed to the interpretation and analysis of the data. Yongpan Li, Yichen Liu, and Cheng-Cheng Liu wrote the manuscript. }
\authordeclaration{The authors declare no conflict of interest.}
\equalauthors{\textsuperscript{1} Yongpan Li and Yichen Liu contributed equally to this work.}
\correspondingauthor{\textsuperscript{2} E-mail:ccliu@bit.edu.cn}


\begin{abstract}
Berry curvature, the imaginary part of the quantum geometry, has provided a microscopic understanding of gyrotropic magneto-optical effects (MOEs), such as Kerr and Faraday effects, yet the microscopic origin of more general MOEs, such as the Voigt effect (or quadratic magneto-optical Kerr effect), remains unresolved. Here, we develop a symmetry-general theory of MOEs in terms of full quantum geometry, and propose the quantum metric, the real part of the quantum geometry, as a novel microscopic mechanism underlying MOEs. The quantum metric and Berry curvature contributions are shown to enter MOEs in a coupled rather than simply additive manner. We identify Berry-curvature-induced MOEs as conventional MOEs and quantum-metric-dominated MOEs as unconventional MOEs. Applying our framework to altermagnets, we find that the quantum metric uncovers previously unanticipated magneto-optical activity in ideal altermagnets without spin-orbit coupling. First-principles calculations for three-dimensional canonical altermagnets FeF$_3$ and MnTe and the emerging two-dimensional twisted bilayer altermagnet CrSBr further demonstrate that unconventional MOEs persist in realistic altermagnets with spin-orbit coupling. Our work develops a full-quantum-geometry-based microscopic theory for MOEs beyond conventional Berry curvature and magnetic order paradigms, and provides a quantum-geometric principle for discovering and controlling magneto-optical responses in altermagnets.
\end{abstract}

\dates{This manuscript was compiled on \today}
\doi{\url{www.pnas.org/cgi/doi/10.1073/pnas.XXXXXXXXXX}}

\maketitle
\thispagestyle{firststyle}
\ifthenelse{\boolean{shortarticle}}{\ifthenelse{\boolean{singlecolumn}}{\abscontentformatted}{\abscontent}}{}

\firstpage[8]{3}

\begin{figure*}[t]
\centering
\includegraphics[width=1\textwidth]{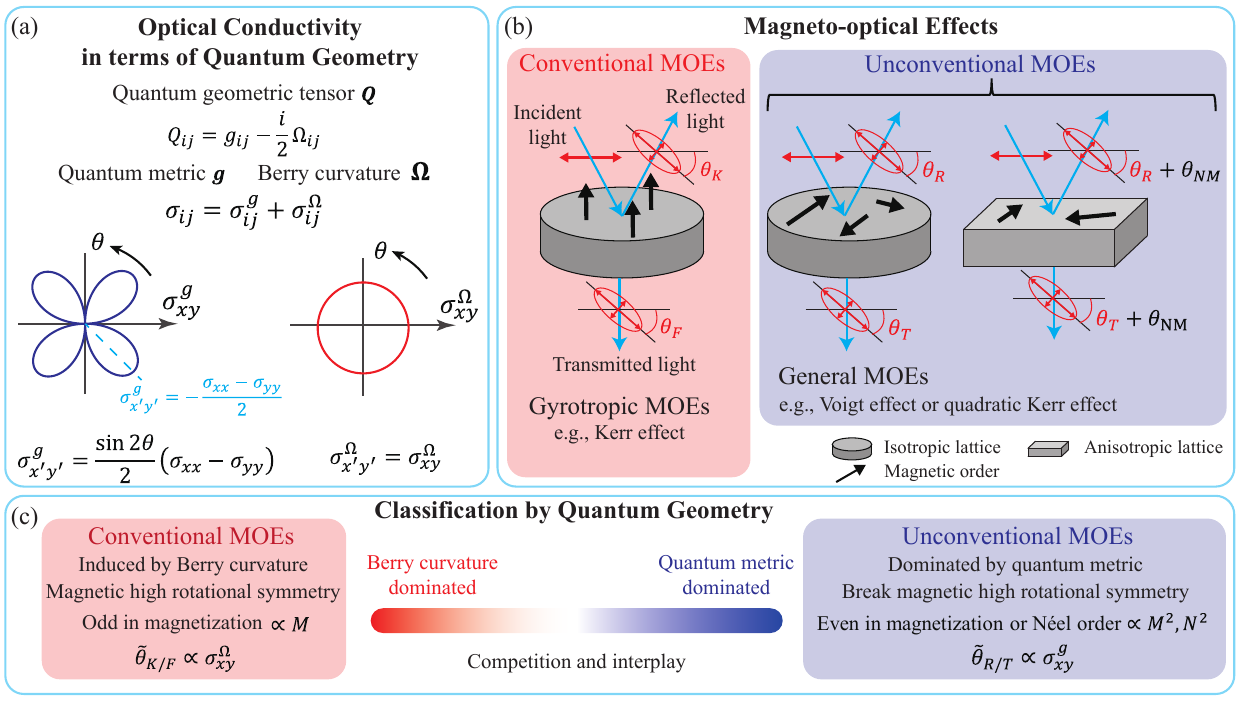}
\caption{Unconventional versus conventional magneto-optical effects (MOEs). (a) The optical conductivity tensor $\boldsymbol{\sigma}$ in terms of full quantum geometry. $\theta$ denotes the rotation angle that transforms the $xy$ coordinate system into the $x'y'$ coordinate system. The transverse optical conductivities contributed by the quantum metric ($\sigma_{xy}^g$) are anisotropic, and those by Berry curvature ($\sigma_{xy}^\Omega$) are isotropic.  (b) Symmetry conditions for conventional and our proposed unconventional MOEs. Conventional MOEs (e.g., Kerr effect) arise when both the lattice and the magnetic configuration are isotropic. Unconventional MOEs (e.g., Voigt effect or quadratic Kerr effect) are present when the lattice is isotropic but the magnetic configuration is not, or when the lattice is anisotropic. When the lattice is anisotropic, the nonmagnetic background $\theta_{\mathrm{NM}}$ should be subtracted. (c) Quantum-geometric classification of conventional and our proposed unconventional MOEs. Conventional MOEs arise exclusively from the Berry curvature, while the quantum metric dominates the contribution to magneto-optical response in unconventional MOEs. As the Berry curvature and quantum metric are time-reversal odd and even, respectively, the intensity of the conventional MOEs scales linearly with magnetization $M$, while that of the unconventional MOEs scales quadratically with $M$ and N\'eel order $N$. In terms of the transverse optical conductivities, the intensity of conventional MOEs is proportional to $\sigma_{xy}^\Omega$, and that of unconventional MOEs is proportional to $\sigma_{xy}^g$.}
\label{fig:moe}
\end{figure*}

\dropcap{Q}uantum geometry provides a unified description of the geometric and topological properties of quantum states~\cite{provost_riemannian_1980,shapere1989geometric,ma_abelian_2010,luhaizhou2024quantume}. The quantum geometric tensor encodes the quantum metric and Berry curvature in its real and imaginary parts, respectively, and provides a powerful framework for understanding a wide range of response phenomena~\cite{nagaosa_anomalous_2010,gao_field_2014,peotta_superfluidity_2015,sodemann_quantum_2015,PhysRevLett.122.227402,torma2022superconductivity,ahn_riemannian_2022,PhysRevLett.132.026301,ezawa2024analytic,PhysRevX.14.011052,yu_quantum_2025}. While the Berry curvature has long been recognized for its physical significance~\cite{berry_quantal_1984,karplus1954hall,xiao_berry_2010,nagaosa_anomalous_2010}, the quantum metric has only emerged as an important physical quantity over the past decade~\cite{peotta_superfluidity_2015,torma2022superconductivity}. In recent years, the quantum-geometric origin of optical conductivity tensor $\boldsymbol{\sigma}$ has begun to be systematically elucidated~\cite{ahn_riemannian_2022}, with both the quantum metric $g$ and Berry curvature $\Omega$ identified as essential contributions to optical conductivity tensor $\boldsymbol{\sigma}\equiv\boldsymbol{\sigma}^g+\boldsymbol{\sigma}^\Omega$ (See Methods and Fig.~\ref{fig:moe}(a)). The emerging understanding of optical conductivity has, in turn, provided new insights into a wide range of phenomena~\cite{komissarov_quantum_2024,verma2025instantaneous,PhysRevB.111.085202,PhysRevB.111.205107}, yet whether and how the full quantum geometry governs magneto-optical effects (MOEs) remains unclear.

MOEs have long been widely used as probes of magnetic properties and symmetry~\cite{PhysRev.186.891,voigt1908magneto,k.h.j.buschow_handbook_1990,kuch_magnetic_2014,PhysRevLett.30.1329,wettling1976magneto,KATO1995713,PhysRevB.51.12633,PhysRevLett.96.167402,tse_giant_2010,feng_large_2015,sivadas_gatecontrollable_2016,huang2017layer,ahn2022theory,PhysRevLett.131.156702,kato2023topological,li2024topological,PhysRevB.111.064428,PhysRevLett.87.047401,nyy3-h8wr,kimak2026ultrafast,10.1063/1.1449436,POSTAVA1997199,PhysRevB.68.064401,wohlrath2025quadratic,PhysRevB.100.064403}. While conventional Berry-curvature-based theories have successfully captured gyrotropic MOEs such as Kerr and Faraday effects~\cite{feng_large_2015,sivadas_gatecontrollable_2016}, more general MOEs such as the Voigt effect~\cite{voigt1908magneto,PhysRevLett.87.047401,nyy3-h8wr,kimak2026ultrafast} (or quadratic magneto-optical Kerr effect~\cite{10.1063/1.1449436,POSTAVA1997199,PhysRevB.68.064401,wohlrath2025quadratic,kimak2026ultrafast,PhysRevB.100.064403}) still lack a unified microscopic formulation in terms of the geometric and topological properties, i.e., quantum geometry, of electronic states. Meanwhile, the connection between MOEs and optical conductivity suggests that the full quantum geometry would be expected to underlie MOEs as well. Yet, conventional Berry-curvature-based theories have exclusively retained Berry-curvature contributions, leaving the role of the quantum metric unclear.

Altermagnets have recently emerged as a rapidly developing class of magnetic materials, attracting broad interest due to the spin-split electronic structures and rich symmetry properties~\cite{PhysRevB.75.115103,PhysRevB.102.014422,gonzalez-hernandez_efficient_2021,ma2021multifunctional,smejkal_conventional_2022,smejkal_emerging_2022,mcclarty_landau_2024,PhysRevX.12.040002}. In the absence of spin-orbit coupling (SOC), ideal altermagnets possess an effective time-reversal symmetry formed by time reversal combined with spin flipping, which enforces a vanishing Berry curvature. With SOC, the diverse magnetic symmetries of altermagnets impose highly selective constraints on the quantum metric and Berry curvature, providing a versatile platform for exploring the role of full quantum geometry in MOEs.

\newpage

Here, we take an important step toward a more complete understanding of MOEs. For the first time, we establish a symmetry-general quantum-geometric theory of MOEs that identifies the quantum metric as a distinct microscopic mechanism beyond the well-established Berry-curvature picture. Our general formulas reveal that quantum metric and Berry curvature enter the magneto-optical response in a coupled rather than simply additive manner. Within our framework, the conventional Berry-curvature-based theory emerges as a special high-symmetry limit: for systems with threefold or higher rotational symmetry, quantum-metric contributions to MOEs are symmetry-forbidden, leaving the Berry curvature as the sole microscopic source. We refer to these Berry-curvature-induced responses as conventional MOEs, and term quantum-metric-dominated responses unconventional MOEs, i.e., cases in which the quantum-metric contribution far exceeds that of the Berry curvature (See Fig.~\ref{fig:moe}(c)). Phenomenologically, unconventional MOEs encompass, but are not limited to, well-known magnetization-even responses such as the Voigt effect (or quadratic magneto-optical Kerr effect), as shown in Fig.~\ref{fig:moe}(b). Through symmetry analysis, we find that many ideal altermagnets exhibit unconventional MOEs despite being magneto-optically inactive within the conventional Berry-curvature picture. Finally, combining our general theory with first-principles calculations, we quantify the roles of the quantum metric and Berry curvature and demonstrate that unconventional MOEs persist in realistic altermagnets with SOC, using the three-dimensional (3D) canonical altermagnets FeF$_3$ and MnTe and the emerging two-dimensional (2D) twisted bilayer altermagnet CrSBr as representative examples.

\section*{Results and Discussion}

\subsection*{Symmetry-general formulas based on full quantum geometry for MOEs}
Consider linearly polarized light incident normally on the surface of materials. Without loss of generality, we take the incident light to propagate along the positive $z$-direction, i.e., $\boldsymbol{E}=E_0e^{i(k_zz-\omega t)}\boldsymbol{x}$. The wave equations in materials are
\begin{equation}
\left(k_0^2-\boldsymbol{k}^2\right)E_i+i\frac{k_0^2}{\omega\epsilon_0}\sigma_{ij}E_j=0,
\end{equation}
where $i,j\in \{x,y\}$ denote the in-plane Cartesian coordinates.

By solving the wave equations, the general formulas for eigenmodes of light in the material, $\boldsymbol{E}_{\pm}$, and the corresponding complex refractive indices, $n_\pm$, read (See Supplementary Information for more details)
\begin{align}\label{eq:eigenmodes}
\boldsymbol{E}_{\pm}=&\left(\frac{1}{2}\pm\frac{\sigma_{xx}-\sigma_{yy}}{2G}\right)E_0e^{i(k_zz-\omega t)}\boldsymbol{x}\nonumber\\
&\pm\frac{\sigma_{xy}^g-\sigma_{xy}^\Omega}{G}E_0e^{i(k_zz-\omega t)}\boldsymbol{y},
\end{align}
\begin{equation}\label{eq:cmplx_refr_ind}
n_\pm=\sqrt{1-\frac{\sigma_{xx}+\sigma_{yy}\pm G}{2i\omega\epsilon_0}},
\end{equation}
where  $G=\sqrt{(\sigma_{yy}-\sigma_{xx})^2+4(\sigma_{xy}^g)^2-4(\sigma_{xy}^\Omega)^2}$. $k_z$ is the effective complex wave vector, i.e., the incident light attenuates as the propagation distance increases. The relation between $k_z$ and $n_\pm$ is given by $k_z=n_\pm k_0$ with $k_0$ the wave propagation number in vacuum, whose dielectric constant is $\epsilon_0$.

To derive general formulas based on full quantum geometry, we reformulate the light-matter eigenmode problem beyond the conventional Berry curvature and circular polarization paradigm. This reformulation is essential because the conventional Berry-curvature-based theories take circularly polarized eigenmodes and therefore cannot accommodate the anisotropic optical conductivity generated by the quantum metric. The resulting symmetry-general reflection complex polarization rotation angle $\widetilde{\theta}_R=\theta_R+i\xi_R$ and the transmission complex polarization rotation angle $\widetilde{\theta}_T=\theta_T+i\xi_T$ naturally incorporate contributions from both the Berry curvature and quantum metric, and are given by
\begin{align}
\widetilde{\theta}_R&=\frac{2(\sigma_{xy}^g-\sigma_{xy}^\Omega)}{\sigma_{xx}-\sigma_{yy}+GR},\label{eq3k}\\
\widetilde{\theta}_T&=\frac{2(\sigma_{xy}^g-\sigma_{xy}^\Omega)}{\sigma_{xx}-\sigma_{yy}+GT}.\label{eqfd}
\end{align}
Here, $R^{-1}$ and $T^{-1}$ are the reflection contrast ratio and transmission contrast ratio of the eigenmodes of light, respectively. For 3D systems, $R$ and $T$ are given by
\begin{equation}
R^{\mathrm{3D}}=\frac{n_+n_--1}{n_+-n_-},\ T^{\mathrm{3D}}=\frac{e^{ik_0n_+l}+e^{ik_0n_-l}}{e^{ik_0n_+l}-e^{ik_0n_-l}},
\end{equation}
where $l$ is the propagation distance of the light. For 2D systems, $R$ and $T$ read
\begin{equation}
R^{\mathrm{2D}}=\frac{r(n_+)+r(n_-)}{r(n_+)-r(n_-)},\ T^{\mathrm{2D}}=\frac{t(n_+)+t(n_-)}{t(n_+)-t(n_-)},\label{eq:T}
\end{equation}
where $t(n_i)$ and $r(n_i)$ are the total complex transmission coefficient and the total complex reflection coefficient, respectively. The explicit expressions for $t(n_i)$ and $r(n_i)$ are given in Supplementary Information.

Equations (\ref{eq:eigenmodes}-\ref{eqfd}) represent the main results of this work. Importantly, Eqs.~(\ref{eq:eigenmodes}-\ref{eqfd}) indicate that the magneto-optical response cannot be decomposed into independent quantum-metric and Berry-curvature contributions: the quantum metric and Berry curvature jointly determine the eigenmodes and the complex refractive indices, and therefore enter the magneto-optical responses nonlinearly.

One should note that, as shown in Fig.~\ref{fig:moe}(b), finite values $\widetilde{\theta}_{R/T}$ of Eqs.~(\ref{eq3k}) and (\ref{eqfd}) can be directly interpreted as a magneto-optical response when $\widetilde{\theta}_{R/T}$ vanish in the nonmagnetic state and become finite in the magnetic state. If the nonmagnetic state already exhibits a finite optical rotation, the nonmagnetic background contribution $\widetilde{\theta}_{\mathrm{NM}}$ needs to be subtracted to isolate the genuine magneto-optical response $\widetilde{\theta}_{R/T}$.

\begin{figure*}
\centering
\includegraphics[width=1\textwidth]{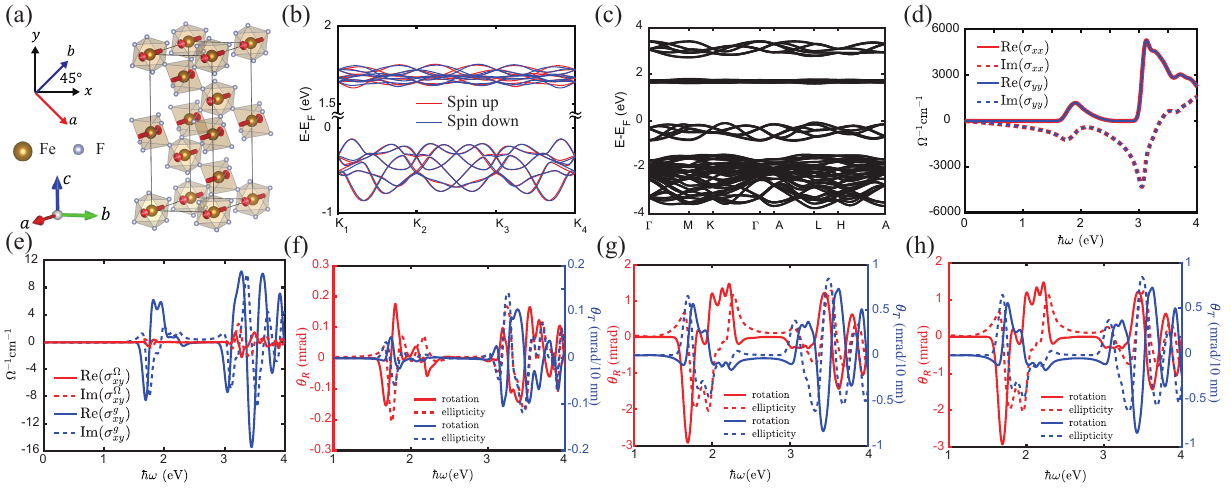}
\caption{The unconventional magneto-optical effects in altermagnets. (a) The magnetic unit cell of 3D FeF$_3$, as drawn using VESTA~\cite{vesta}. The polarization direction of the incident light lies within the $ab$ plane, forming an angle of 45\degree\ with respect to the $a$-axis. (b) The DFT spin-splitting band structure without SOC. The positions of the four momenta are given by $\mathrm{K}_1=(1/3,1/3,1/4)$, $\mathrm{K}_2=(1/3,-2/3,1/4)$, $\mathrm{K}_3=(1/3,1/3,-1/4)$, $\mathrm{K}_4=(1/3,-2/3,-1/4)$. (c) The DFT band structure with SOC. (d) The longitudinal optical conductivities $\sigma_{xx}$ and $\sigma_{yy}$. (e) The transverse conductivities $\sigma_{xy}\equiv\sigma_{xy}^g+\sigma_{xy}^\Omega$. (f-h) Calculated reflection and specific transmission polarization rotation angles, i.e., $\theta_R$ and $\theta_T$,  for the cases where the transverse optical conductivities are contributed exclusively by (f) the Berry curvature, (g) the quantum metric, and (h) both. The calculations in (d)--(h) have all incorporated SOC. Results for 3D altermagnet MnTe and 2D twisted bilayer altermagnet CrSBr are provided in the Supplementary Information.}
\label{fig:FeF3}
\end{figure*}

\subsection*{Conventional Berry curvature mechanism}
We briefly review the gyrotropic MOEs and demonstrate that the conventional Berry-curvature-based theories are naturally recovered from our symmetry-general quantum-geometric framework under specific symmetry constraints. The conventional Berry-curvature-based theories require the magnetic system to retain high rotational symmetry (threefold or higher rotational symmetry). Under high rotational symmetry, the longitudinal optical conductivities satisfy $\sigma_{xx}=\sigma_{yy}$, while the transverse optical conductivities are antisymmetric ($\sigma_{xy}=-\sigma_{yx}$). Since the quantum metric is symmetric, the quantum-metric contribution ($\sigma_{xy}^g=\sigma_{yx}^g=0$) to the transverse conductivity is forbidden by symmetry. The Berry-curvature contribution ($\sigma_{xy}^\Omega=-\sigma_{yx}^\Omega\neq0$) survives as the Berry curvature is antisymmetric.

Substituting the above symmetry-constrained antisymmetric optical conductivity tensor into Eq.~(\ref{eq3k}) and Eq.~(\ref{eqfd}), our general expressions reduce to the well-known and commonly used 3D Kerr and Faraday angle formulas~\cite{PhysRev.186.891,wettling1976magneto,PhysRevLett.96.167402,RevModPhys.82.2731,feng_large_2015} (See Supplementary Information for more details)
\begin{align}
\widetilde{\theta}_K=&-\frac{\sigma_{xy}^\Omega}{\sigma_{xx}\sqrt{1+i\sigma_{xx}/\omega\epsilon_0}},\\
\widetilde{\theta}_F=&\frac{k_0 l}{2\omega\epsilon_0}\frac{\sigma_{xy}^\Omega}{\sqrt{1+i\sigma_{xx}/\omega\epsilon_0}},
\end{align}
where the standard approximation $|\sigma_{xx}|\gg|\sigma_{xy}^\Omega|$ has been used. Therefore, the conventional Berry-curvature-based theories correspond to special high-symmetry limits of our symmetry-general quantum-geometric formulation.

\subsection*{Novel quantum metric mechanism}
Inspection of Eqs.~(\ref{eq3k}) and (\ref{eqfd}) leads to a surprising conclusion: a finite magneto-optical response can arise even in the complete absence of Berry-curvature contribution. Specifically, when the Berry-curvature-induced transverse optical conductivity vanishes, i.e., $\sigma_{xy}^\Omega=-\sigma_{yx}^\Omega=0$, the magneto-optical responses remain finite as long as the quantum-metric contribution $\sigma_{xy}^g=\sigma_{yx}^g\neq0$. There is a fundamentally distinct microscopic mechanism of MOEs beyond the conventional Berry-curvature picture.

As the high rotational symmetry suppresses the quantum-metric contribution to MOEs, the quantum-metric mechanism becomes active only when high rotational symmetry is broken. The symmetry breaking can originate either from magnetic order or from lattice anisotropy. For crystal structures with high rotational symmetry, magnetic ordering can lower the rotational symmetry through either the orientations of magnetic moments or their spatial arrangement among magnetic sublattices, and unlock the quantum metric mechanism. In this case, when the magnetic order is ferromagnetic or antiferromagnetic, the magneto-optical responses correspond to the Voigt effect or quadratic magneto-optical Kerr effects. Alternatively, in anisotropic crystals where high rotational symmetry is already absent, the quantum-metric contribution is always symmetry-allowed. In this case, the genuine magneto-optical response should be obtained by subtracting the nonmagnetic optical anisotropy background from the total optical response.

In recent years, particularly with the emergence of altermagnets, a growing number of magnetic systems with low magnetic symmetry have been discovered, where the quantum metric mechanism can play an increasingly important role in determining magneto-optical responses. Altermagnets provide an ideal platform for exploring the quantum metric mechanism and for exploring the competition between the quantum metric mechanism and the Berry curvature mechanism.

An ideal altermagnet is a collinear and crystal-symmetry-enforced fully compensated magnet described by spin symmetry, whose N\'eel vector is odd under some real-space symmetries other than inversion and translation symmetries~\cite{PhysRevB.102.014422,gonzalez-hernandez_efficient_2021,ma2021multifunctional,smejkal_conventional_2022,smejkal_emerging_2022,PhysRevX.12.021016,guo2023quantum, mcclarty_landau_2024,bailing2024altermag}. Due to the decoupling of the real-space symmetry and spin-space symmetry, ideal altermagnets always possess an effective time-reversal symmetry, i.e., a time-reversal symmetry combined with a spin inversion symmetry, which enforces a vanishing Berry-curvature contribution. However, there are many 2D altermagnets without high rotational symmetry and thus show unconventional MOEs. For 3D systems, even if an altermagnet possesses high rotational symmetry, not all of its cleavage planes have high rotational symmetry. As long as the light is incident on a surface without high rotational symmetry, the quantum-metric-induced MOEs are still present. Even when the crystal surface possesses high rotational symmetry, the spatial arrangement of magnetic moments can break this symmetry, allowing quantum-metric-induced MOEs.

In real altermagnets, although SOC may be very small, it is always present, coupling the real-space symmetry and spin-space symmetry, thus preventing the existence of effective time-reversal symmetry. Even though altermagnets with SOC are capable of exhibiting conventional MOEs, we demonstrate in the following sections that unconventional MOEs, i.e., the quantum-metric-dominated MOEs, can frequently emerge in real altermagnets.

\subsection*{Unconventional MOEs in altermagnetic materials}

As an application of our full-quantum-geometry-based theory of MOEs, we perform first-principles calculations for three representative altermagnets, where the SOC is included and the symmetry description becomes magnetic group symmetry. We present the 3D altermagnet FeF$_3$ here because its isotropic crystal structure avoids additional nonmagnetic optical anisotropy backgrounds. The 3D altermagnet MnTe and the 2D twisted bilayer altermagnet CrSBr are shown in the Supplementary Information. Because of their anisotropic lattice or low-symmetry cleavage plane, the total optical response contains a nonmagnetic optical-anisotropy background, which must be subtracted to isolate the magnetic contribution following the standard procedure commonly adopted in the literature~\cite{PhysRevLett.116.097204,nanolett1c01719,6gbq-d7jk}.

The iron trifluoride FeF$_3$ has a G-type antiferromagnetic structure and belongs to the space group $R\overline{3}c$ (No. 167) [Fig.~\ref{fig:FeF3}(a)]. The magnetic structure of FeF$_3$ has been confirmed by neutron diffraction experiments with a high N{\'e}el temperature of 394 K~\cite{wollan1958antiferromagnetic,LEE201894}. The magnetic sublattices of FeF$_3$ are connected by $C_{2a}$ and, FeF$_3$ exhibits an anisotropic $g$-wave spin-splitting~\cite{smejkal_conventional_2022} as shown in Fig. \ref{fig:FeF3}(b). Although the inversion symmetry is present, it connects magnetic atoms with the same magnetic moments. The magnetic space group of FeF$_3$ is $C2'/2'$ (No. 15.89), and the band structure with SOC is shown in Fig. \ref{fig:FeF3}(c).

As illustrated in the inset of Fig. \ref{fig:FeF3}(a), we choose the incident linearly polarized light to propagate along the $\boldsymbol{c}$ direction, and define the $\boldsymbol{x}$ and $\boldsymbol{y}$ axes in the $\boldsymbol{ab}$ plane such that the $\boldsymbol{x}$ axis makes a 45-degree angle with the $\boldsymbol{a}$ axis. In Fig. \ref{fig:FeF3}(d) and (e), the longitudinal and transverse optical conductivities are shown in the range of 0--4 eV. In Fig. \ref{fig:FeF3}, we present the respective contributions of the quantum metric ($\sigma_{xy}^g$) and the Berry curvature ($\sigma_{xy}^\Omega$) to the transverse optical conductivity $\sigma_{xy}$. One can find that the contributions of the quantum metric to the transverse optical conductivity are significantly greater than those of the Berry curvature.

Based on our quantum-geometry-based MOE theory Eqs. (\ref{eq3k}, \ref{eqfd}), the reflection polarization rotation angle and specific transmission polarization rotation angle, i.e., transmission polarization rotation angle per unit thickness $l$ with $l$=10 nm, are calculated in the range of 1--4 eV considering the respective contribution of the Berry curvature [Fig. \ref{fig:FeF3}(f)] and the quantum metric [Fig. \ref{fig:FeF3}(g)] to the transverse optical conductivities. The maximum reflection rotation angle and ellipticity are 0.17 mrad and -0.20 mrad, when considering exclusively the Berry curvature's contribution to the transverse conductivities, and surge to -2.9 mrad and -2.0 mrad, when considering exclusively the quantum metric's contribution. The maximum specific transmission rotation angle and ellipticity are 0.11 mrad/10 nm and 0.14 mrad/10 nm, when considering exclusively the Berry curvature's contribution to the transverse conductivities, and are significantly increased to -0.84 mrad/10 nm and 0.84 mrad/10 nm, when considering exclusively the quantum metric's contribution. In Fig. \ref{fig:FeF3}(h), we calculate the total reflection polarization rotation angle and specific transmission polarization rotation angle considering both the contributions of the quantum metric and the Berry curvature. Since the quantum metric dominates the contribution to MOEs, we find that the total reflection polarization rotation angle and total specific transmission polarization rotation angle are almost identical to those calculated without considering Berry curvature.

Our calculations on FeF$_3$ illustrate the physical origin of unconventional MOEs. In the absence of SOC, in FeF$_3$, the effective time-reversal symmetry prohibits the contribution of Berry curvature, while the three-fold rotational symmetry prohibits the contribution of the quantum metric. Therefore, both contributions in FeF$_3$ depend on the breaking of symmetry induced by SOC and by the orientations of the magnetic moments on the magnetic atoms. SOC does not merely activate the conventional Berry-curvature channel. More importantly, by locking the magnetic moments to the lattice, SOC lowers the relevant magnetic rotational symmetry and unlocks a much larger quantum-metric channel.

\subsection*{Discussion}

In traditional Berry-curvature-based theory, the polarization direction of the incident light---i.e., the orientation of its electric field---does not affect the magnitude of the MOE signal. By contrast, in FeF$_3$, we find that the MOE signal depends strongly on the polarization direction of the incident light (with the $x$ direction taken as the polarization axis in this work). This difference originates from the symmetry of the plane of incidence. In the traditional framework, the plane of incidence has high rotational symmetry, and the eigenmodes of light are left- and right-circularly polarized. Decomposing incident linearly polarized light into these circular eigenmodes always yields equal amplitudes, independent of the polarization direction. As a result, the Kerr and Faraday angles are the same for different linear polarization directions in the traditional theory. In the more general case where high rotational symmetry is absent, however, the eigenmodes become elliptically polarized [see Eq. (\ref{eq:eigenmodes})]. Decomposing linearly polarized light with different polarization directions into these elliptically polarized eigenmodes produces direction-dependent amplitude ratios. Consequently, linearly polarized light with different polarization directions gives rise to different reflection or transmission polarization rotation angles in the unconventional MOEs we propose. A pronounced dependence of the magneto-optical response on the polarization direction can serve as an experimental signature distinguishing unconventional MOEs from conventional MOEs.

In this work, we challenge the prevailing microscopic quantum-geometric understanding in which MOEs are associated exclusively with the Berry curvature. We show that the standard formulas for what we call conventional MOEs apply only to a limited subset of altermagnets with high rotational symmetry. We identify a new mechanism---quantum-metric-induced or -dominated responses---that both deepens and broadens the conceptual framework of MOEs. Because our predictions can be tested using standard MOE experimental techniques, they provide a direct route to experimental validation. More broadly, our results highlight altermagnets as a promising platform for quantum-geometry-driven optical phenomena, with important implications for future magneto-optical applications and for stimulating further theoretical and experimental exploration.

\acknow{The work is supported by the NSF of China (Grant No. 12374055), the Science Fund for Creative Research Groups of NSFC (Grant No. 12321004), and the National Key R\&D Program of China (Grant No. 2020YFA0308800).}

\showacknow{} 

\bibsplit[12]

\bibliography{reference}

@book{voigt1908magneto,
  title={Magneto-und elektrooptik},
  author={Voigt, Woldemar},
  number={3},
  year={1908},
  publisher={BG Teubner}
}

@article{PhysRev.186.891,
  title = {{Ultraviolet Magneto-Optical Properties of Single-Crystal Orthoferrites, Garnets, and Other Ferric Oxide Compounds}},
  author = {Kahn, Frederic J. and Pershan, P. S. and Remeika, J. P.},
  journal = {Phys. Rev.},
  volume = {186},
  issue = {3},
  pages = {891--918},
  numpages = {0},
  year = {1969},
  month = {Oct},
  publisher = {American Physical Society},
  doi = {10.1103/PhysRev.186.891},
}

@article{vesta,
author = "Momma, Koichi and Izumi, Fujio",
title = "{{\it VESTA3} for three-dimensional visualization of crystal, volumetric and morphology data}",
journal = "J. Appl. Crystallogr.",
year = "2011",
volume = "44",
number = "6",
pages = "1272--1276",
month = "Dec",
doi = {10.1107/S0021889811038970},
}

@article{aversa_nonlinear_1995,
  title = {{Nonlinear Optical Susceptibilities of Semiconductors: Results with a Length-Gauge Analysis}},
  author = {Aversa, Claudio and Sipe, J. E.},
  year = {1995},
  journal = {Phys. Rev. B},
  volume = {52},
  number = {20},
  pages = {14636--14645},
  doi = {10.1103/PhysRevB.52.14636}
}

@article{ahn_riemannian_2022,
  title = {Riemannian Geometry of Resonant Optical Responses},
  author = {Ahn, Junyeong and Guo, Guang-Yu and Nagaosa, Naoto and Vishwanath, Ashvin},
  year = {2022},
  journal = {Nat. Phys.},
  volume = {18},
  number = {3},
  pages = {290--295},
  doi = {10.1038/s41567-021-01465-z}
}

@article{ma_abelian_2010,
  title = {Abelian and Non-{{Abelian}} Quantum Geometric Tensor},
  author = {Ma, Yu-Quan and Chen, Shu and Fan, Heng and Liu, Wu-Ming},
  year = {2010},
  journal = {Phys. Rev. B},
  volume = {81},
  number = {24},
  pages = {245129},
  doi = {10.1103/PhysRevB.81.245129}
}

@article{verma2025instantaneous,
  title={Instantaneous response and quantum geometry of insulators},
  author={Verma, Nishchhal and Queiroz, Raquel},
  journal={Proc. Natl. Acad. Sci. U.S.A.},
  volume={122},
  number={49},
  pages={e2405837122},
  year={2025},
  publisher={National Academy of Sciences},
  doi={10.1073/pnas.2405837122}
}

@article{gonzalez-hernandez_efficient_2021,
  title = {Efficient {{Electrical Spin Splitter Based}} on {{Nonrelativistic Collinear Antiferromagnetism}}},
  author = {{Gonz{\'a}lez-Hern{\'a}ndez}, Rafael and {\v S}mejkal, Libor and V{\'y}born{\'y}, Karel and Yahagi, Yuta and Sinova, Jairo and Jungwirth, Tom{\'a}{\v s} and {\v Z}elezn{\'y}, Jakub},
  year = {2021},
  journal = {Phys. Rev. Lett.},
  volume = {126},
  number = {12},
  pages = {127701},
  doi = {10.1103/PhysRevLett.126.127701}
}

@article{mcclarty_landau_2024,
  title = {Landau {{Theory}} of {{Altermagnetism}}},
  author = {McClarty, Paul A. and Rau, Jeffrey G.},
  year = {2024},
  journal = {Phys. Rev. Lett.},
  volume = {132},
  number = {17},
  pages = {176702},
  doi = {10.1103/PhysRevLett.132.176702}
}

@article{smejkal_conventional_2022,
  title = {Beyond {{Conventional Ferromagnetism}} and {{Antiferromagnetism}}: {{A Phase}} with {{Nonrelativistic Spin}} and {{Crystal Rotation Symmetry}}},
  author = {{\v S}mejkal, Libor and Sinova, Jairo and Jungwirth, Tomas},
  year = {2022},
  journal = {Phys. Rev. X},
  volume = {12},
  number = {3},
  pages = {031042},
  doi = {10.1103/PhysRevX.12.031042}
}

@article{smejkal_emerging_2022,
  title = {Emerging {{Research Landscape}} of {{Altermagnetism}}},
  author = {{\v S}mejkal, Libor and Sinova, Jairo and Jungwirth, Tomas},
  year = {2022},
  journal = {Phys. Rev. X},
  volume = {12},
  number = {4},
  pages = {040501},
  doi = {10.1103/PhysRevX.12.040501}
}

@book{k.h.j.buschow_handbook_1990,
	title = {Handbook of {Ferromagnetic} {Materials}},
	publisher = {North Holland},
	author = {{K.H.J. Buschow} and {E.P. Wohlfarth}},
	year = {1990},
}

@book{kuch_magnetic_2014,
  title = {Magnetic {{Microscopy}} of {{Layered Structures}}},
  author = {Kuch, Wolfgang and Sch{\"a}fer, Rudolf and Fischer, Peter and Hillebrecht, F. U.},
  year = {2014},
  publisher = {Springer Berlin, Heidelberg},
  isbn = {978-3-662-44532-7},
}

@article{provost_riemannian_1980,
  title = {Riemannian Structure on Manifolds of Quantum States},
  author = {Provost, J. P. and Vallee, G.},
  year = {1980},
  journal = {Commun.Math. Phys.},
  volume = {76},
  number = {3},
  pages = {289--301},
  doi = {10.1007/BF02193559}
}

@book{shapere1989geometric,
  title={Geometric phases in physics},
  author={Shapere, Alfred and Wilczek, Frank},
  volume={5},
  year={1989},
  publisher={World scientific}
}

@article{sodemann_quantum_2015,
  title = {Quantum {{Nonlinear Hall Effect Induced}} by {{Berry Curvature Dipole}} in {{Time-Reversal Invariant Materials}}},
  author = {Sodemann, Inti and Fu, Liang},
  year = {2015},
  journal = {Phys. Rev. Lett.},
  volume = {115},
  number = {21},
  pages = {216806},
  doi = {10.1103/PhysRevLett.115.216806}
}

@article{gao_field_2014,
  title = {Field {{Induced Positional Shift}} of {{Bloch Electrons}} and {{Its Dynamical Implications}}},
  author = {Gao, Yang and Yang, Shengyuan A. and Niu, Qian},
  year = {2014},
  journal = {Phys. Rev. Lett.},
  volume = {112},
  number = {16},
  pages = {166601},
  doi = {10.1103/PhysRevLett.112.166601}
}

@article{PhysRevLett.132.026301,
  title = {Unification of Nonlinear Anomalous Hall Effect and Nonreciprocal Magnetoresistance in Metals by the Quantum Geometry},
  author = {Kaplan, Daniel and Holder, Tobias and Yan, Binghai},
  journal = {Phys. Rev. Lett.},
  volume = {132},
  issue = {2},
  pages = {026301},
  numpages = {7},
  year = {2024},
  doi = {10.1103/PhysRevLett.132.026301},
}

@article{luhaizhou2024quantume,
    author = {Liu, Tianyu and Qiang, Xiao-Bin and Lu, Hai-Zhou and Xie, X C},
    title = {Quantum geometry in condensed matter},
    journal = {Natl. Sci. Rev.},
    pages = {nwae334},
    year = {2024},
    month = {09},
    issn = {2095-5138},
    doi = {10.1093/nsr/nwae334},
}

@article{torma2022superconductivity,
  title={Superconductivity, superfluidity and quantum geometry in twisted multilayer systems},
  author={T{\"o}rm{\"a}, P{\"a}ivi and Peotta, Sebastiano and Bernevig, Bogdan A},
  journal={Nat. Rev. Phys.},
  volume={4},
  number={8},
  pages={528--542},
  year={2022},
  doi={10.1038/s42254-022-00466-y}
}

@article{ezawa2024analytic,
  title = {Analytic approach to quantum metric and optical conductivity in Dirac models with parabolic mass in arbitrary dimensions},
  author = {Ezawa, Motohiko},
  journal = {Phys. Rev. B},
  volume = {110},
  issue = {19},
  pages = {195437},
  numpages = {11},
  year = {2024},
  month = {Nov},
  publisher = {American Physical Society},
  doi = {10.1103/PhysRevB.110.195437}
}

@article{PhysRevX.14.011052,
  title = {Fundamental Bound on Topological Gap},
  author = {Onishi, Yugo and Fu, Liang},
  journal = {Phys. Rev. X},
  volume = {14},
  issue = {1},
  pages = {011052},
  numpages = {12},
  year = {2024},
  month = {Mar},
  doi = {10.1103/PhysRevX.14.011052},
}

@article{komissarov_quantum_2024,
  title = {The Quantum Geometric Origin of Capacitance in Insulators},
  author = {Komissarov, Ilia and Holder, Tobias and Queiroz, Raquel},
  year = {2024},
  journal = {Nat. Commun.},
  volume = {15},
  number = {1},
  pages = {4621},
  publisher = {Nature Publishing Group},
  doi = {10.1038/s41467-024-48808-x}
}

@article{sivadas_gatecontrollable_2016,
  title = {Gate-{{Controllable Magneto-optic Kerr Effect}} in {{Layered Collinear Antiferromagnets}}},
  author = {Sivadas, Nikhil and Okamoto, Satoshi and Xiao, Di},
  year = {2016},
  journal = {Phys. Rev. Lett.},
  volume = {117},
  number = {26},
  pages = {267203},
  doi = {10.1103/PhysRevLett.117.267203}
}

@article{ma2021multifunctional,
  title={Multifunctional antiferromagnetic materials with giant piezomagnetism and noncollinear spin current},
  author={Ma, Hai-Yang and Hu, Mengli and Li, Nana and Liu, Jianpeng and Yao, Wang and Jia, Jin-Feng and Liu, Junwei},
  journal={Nat. Commun.},
  volume={12},
  number={1},
  pages={2846},
  year={2021},
  publisher={Nature Publishing Group UK London},
  doi={10.1038/s41467-021-23127-7}
}

@article{PhysRevB.102.014422,
  title = {Giant momentum-dependent spin splitting in centrosymmetric low-$Z$ antiferromagnets},
  author = {Yuan, Lin-Ding and Wang, Zhi and Luo, Jun-Wei and Rashba, Emmanuel I. and Zunger, Alex},
  journal = {Phys. Rev. B},
  volume = {102},
  issue = {1},
  pages = {014422},
  numpages = {13},
  year = {2020},
  month = {Jul},
  publisher = {American Physical Society},
  doi = {10.1103/PhysRevB.102.014422},
}

@article{feng_large_2015,
  title = {Large Magneto-Optical {{Kerr}} Effect in Noncollinear Antiferromagnets $\mathrm{Mn}_{3}{X}({X}=\mathrm{Rh},\mathrm{Ir},\mathrm{Pt})$},
  author = {Feng, Wanxiang and Guo, Guang-Yu and Zhou, Jian and Yao, Yugui and Niu, Qian},
  year = {2015},
  month = oct,
  journal = {Phys. Rev. B},
  volume = {92},
  number = {14},
  pages = {144426},
  publisher = {American Physical Society},
  doi = {10.1103/PhysRevB.92.144426}
}

@article{huang2017layer,
  title={Layer-dependent ferromagnetism in a van der Waals crystal down to the monolayer limit},
  author={Huang, Bevin and Clark, Genevieve and Navarro-Moratalla, Efr{\'e}n and Klein, Dahlia R and Cheng, Ran and Seyler, Kyle L and Zhong, Ding and Schmidgall, Emma and McGuire, Michael A and Cobden, David H and others},
  journal={Nature},
  volume={546},
  number={7657},
  pages={270--273},
  year={2017},
  doi={10.1038/nature22391}
}

@article{PhysRevLett.30.1329,
  title = {{Magneto-optic Kerr Effect in Ni, Co, and Fe}},
  author = {Erskine, J. L. and Stern, E. A.},
  journal = {Phys. Rev. Lett.},
  volume = {30},
  issue = {26},
  pages = {1329--1332},
  numpages = {0},
  year = {1973},
  month = {Jun},
  publisher = {American Physical Society},
  doi = {10.1103/PhysRevLett.30.1329},
}

@article{KATO1995713,
title = {{Magneto-optical effect in MnPt$_3$ alloy films}},
journal = {J. Magn. and Magn. Mater.},
volume = {140-144},
pages = {713-714},
year = {1995},
issn = {0304-8853},
doi = {10.1016/0304-8853(94)01572-4},
author = {T. Kato and H. Kikuzawa and S. Iwata and S. Tsunashima and S. Uchiyama},
}

@article{PhysRevB.51.12633,
  title = {{Band-theoretical investigation of the magneto-optical Kerr effect in Fe and Co multilayers}},
  author = {Guo, G. Y. and Ebert, H.},
  journal = {Phys. Rev. B},
  volume = {51},
  issue = {18},
  pages = {12633--12643},
  numpages = {0},
  year = {1995},
  month = {May},
  publisher = {American Physical Society},
  doi = {10.1103/PhysRevB.51.12633},
}

@article{li2024topological,
  title={{Topological Kerr effects in two-dimensional magnets with broken inversion symmetry}},
  author={Li, Xiaoyin and Liu, Caixing and Zhang, Ying and Zhang, Shunhong and Zhang, Huisheng and Zhang, Yuchen and Meng, Wenjie and Hou, De and Li, Tao and Kang, Chaoyang and others},
  journal={Nat. Phys.},
  volume = {20},
  pages = {1145--1151},
  year={2024},
  doi={10.1038/s41567-024-02465-5}
}

@article{tse_giant_2010,
  title = {Giant {{Magneto-Optical Kerr Effect}} and {{Universal Faraday Effect}} in {{Thin-Film Topological Insulators}}},
  author = {Tse, Wang-Kong and MacDonald, A. H.},
  year = {2010},
  journal = {Phys. Rev. Lett.},
  volume = {105},
  number = {5},
  pages = {057401},
  doi = {10.1103/PhysRevLett.105.057401}
}

@article{PhysRevLett.96.167402,
  title = {{Magneto-Optical Kerr Effects of Yttrium-Iron Garnet Thin Films Incorporating Gold Nanoparticles}},
  author = {Tomita, Satoshi and Kato, Takeshi and Tsunashima, Shigeru and Iwata, Satoshi and Fujii, Minoru and Hayashi, Shinji},
  journal = {Phys. Rev. Lett.},
  volume = {96},
  issue = {16},
  pages = {167402},
  numpages = {4},
  year = {2006},
  month = {Apr},
  publisher = {American Physical Society},
  doi = {10.1103/PhysRevLett.96.167402},
}

@article{wettling1976magneto,
  title={Magneto-optics of ferrites},
  author={Wettling, W},
  journal={J. Magn. Magn. Mater.},
  volume={3},
  number={1-2},
  pages={147--160},
  year={1976},
  publisher={Elsevier},
  doi={10.1016/0304-8853(76)90026-3}
}

@article{ahn2022theory,
  title={{Theory of optical axion electrodynamics and application to the Kerr effect in topological antiferromagnets}},
  author={Ahn, Junyeong and Xu, Su-Yang and Vishwanath, Ashvin},
  journal={Nat. Commun.},
  volume={13},
  number={1},
  pages={7615},
  year={2022},
  doi={10.1038/s41467-022-35248-8}
}

@article{PhysRevLett.131.156702,
  title = {{Magneto-Optical Detection of the Orbital Hall Effect in Chromium}},
  author = {Lyalin, Igor and Alikhah, Sanaz and Berritta, Marco and Oppeneer, Peter M. and Kawakami, Roland K.},
  journal = {Phys. Rev. Lett.},
  volume = {131},
  issue = {15},
  pages = {156702},
  numpages = {6},
  year = {2023},
  month = {Oct},
  publisher = {American Physical Society},
  doi = {10.1103/PhysRevLett.131.156702}
}

@article{kato2023topological,
  title={{Topological magneto-optical effect from skyrmion lattice}},
  author={Kato, Yoshihiro D and Okamura, Yoshihiro and Hirschberger, Max and Tokura, Yoshinori and Takahashi, Youtarou},
  journal={Nat. Commun.},
  volume={14},
  number={1},
  pages={5416},
  year={2023},
  doi={10.1038/s41467-023-41203-y}
}

@article{VASP,
  title={Efficient iterative schemes for $ab$ initio total-energy calculations using a plane-wave basis set},
  author={Kresse, Georg and Furthm{\"u}ller, J{\"u}rgen},
  journal={Phys. Rev. B},
  volume={54},
  number={16},
  pages={11169},
  year={1996},
  publisher={APS},
  doi = {10.1103/physrevb.54.11169}
}

@article{wannier90,
  title={wannier90: A tool for obtaining maximally-localised Wannier functions},
  author={Mostofi, Arash A and Yates, Jonathan R and Lee, Young-Su and Souza, Ivo and Vanderbilt, David and Marzari, Nicola},
  journal={Comput. Phys. Commun.},
  volume={178},
  number={9},
  pages={685--699},
  year={2008},
  publisher={Elsevier},
  doi = {10.1016/j.cpc.2007.11.016}
}

@article{PBE,
  title={Generalized gradient approximation made simple},
  author={Perdew, John P and Burke, Kieron and Ernzerhof, Matthias},
  journal={Phys. Rev. Lett.},
  volume={77},
  number={18},
  pages={3865},
  year={1996},
  publisher={APS},
  doi = {10.1103/physrevlett.77.3865}
}

@article{RevModPhys.82.2731,
  title = {Ultrafast optical manipulation of magnetic order},
  author = {Kirilyuk, Andrei and Kimel, Alexey V. and Rasing, Theo},
  journal = {Rev. Mod. Phys.},
  volume = {82},
  issue = {3},
  pages = {2731--2784},
  numpages = {0},
  year = {2010},
  month = {Sep},
  publisher = {American Physical Society},
  doi = {10.1103/RevModPhys.82.2731},
}

@article{bailing2024altermag,
author = {Bai, Ling and Feng, Wanxiang and Liu, Siyuan and {\v{S}}mejkal, Libor and Mokrousov, Yuriy and Yao, Yugui},
title = {Altermagnetism: Exploring New Frontiers in Magnetism and Spintronics},
journal = {Adv. Funct. Mater.},
volume = {34},
number = {49},
pages = {2409327},
doi = {10.1002/adfm.202409327},
year = {2024}
}

@article{PhysRevB.111.064428,
  title = {Magneto-optical conductivity and circular dichroism in $d$-wave altermagnets},
  author = {Chen, Wang and Zhou, Xiaoying and Lou, Wen-Kai and Chang, Kai},
  journal = {Phys. Rev. B},
  volume = {111},
  issue = {6},
  pages = {064428},
  numpages = {13},
  year = {2025},
  month = {Feb},
  publisher = {American Physical Society},
  doi = {10.1103/PhysRevB.111.064428},
  url = {https://link.aps.org/doi/10.1103/PhysRevB.111.064428}
}

@article{PhysRevB.75.115103,
  title = {Fermi liquid instabilities in the spin channel},
  author = {Wu, Congjun and Sun, Kai and Fradkin, Eduardo and Zhang, Shou-Cheng},
  journal = {Phys. Rev. B},
  volume = {75},
  issue = {11},
  pages = {115103},
  numpages = {25},
  year = {2007},
  month = {Mar},
  publisher = {American Physical Society},
  doi = {10.1103/PhysRevB.75.115103},
  url = {https://link.aps.org/doi/10.1103/PhysRevB.75.115103}
}

@article{PhysRevX.12.021016,
  title = {Spin-Group Symmetry in Magnetic Materials with Negligible Spin-Orbit Coupling},
  author = {Liu, Pengfei and Li, Jiayu and Han, Jingzhi and Wan, Xiangang and Liu, Qihang},
  journal = {Phys. Rev. X},
  volume = {12},
  issue = {2},
  pages = {021016},
  numpages = {19},
  year = {2022},
  month = {Apr},
  publisher = {American Physical Society},
  doi = {10.1103/PhysRevX.12.021016},
}

@article{guo2023quantum,
  title={Quantum anomalous hall effect in collinear antiferromagnetism},
  author={Guo, Peng-Jie and Liu, Zheng-Xin and Lu, Zhong-Yi},
  journal={npj Computational Materials},
  volume={9},
  number={1},
  pages={70},
  year={2023},
  doi={10.1038/s41524-023-01025-4}
}

@article{PhysRevX.12.040002,
  title = {Editorial: Altermagnetism---A New Punch Line of Fundamental Magnetism},
  author = {Mazin, Igor},
  collaboration = {The PRX Editors},
  journal = {Phys. Rev. X},
  volume = {12},
  issue = {4},
  pages = {040002},
  numpages = {3},
  year = {2022},
  month = {Dec},
  publisher = {American Physical Society},
  doi = {10.1103/PhysRevX.12.040002},
}

@article{LEE201894,
title = {Weak-ferromagnetism of CoF3 and FeF3},
journal = {Physica B: Condensed Matter},
volume = {551},
pages = {94-97},
year = {2018},
doi = {https://doi.org/10.1016/j.physb.2017.11.082},
author = {Sanghyun Lee and Shuki Torii and Yoshihisa Ishikawa and Masao Yonemura and Taketo Moyoshi and Takashi Kamiyama}
}

@article{wollan1958antiferromagnetic,
  title={Antiferromagnetic properties of the iron group trifluorides},
  author={Wollan, EO and Child, HR and Koehler, WC and Wilkinson, MK},
  journal={Physical Review},
  volume={112},
  number={4},
  pages={1132},
  year={1958},
  publisher={APS},
  doi={10.1103/PhysRev.112.1132}
}

@article{hu2025anomalous,
  title={Anomalous coherence length in superconductors with quantum metric},
  author={Hu, Jin-Xin and Chen, Shuai A and Law, Kam Tuen},
  journal={Communications Physics},
  volume={8},
  number={1},
  pages={20},
  year={2025},
  doi={10.1038/s42005-024-01930-0},
  publisher={Nature Publishing Group UK London}
}

@article{yu_quantum_2025,
  title = {Quantum Geometry in Quantum Materials},
  author = {Yu, Jiabin and Bernevig, B. Andrei and Queiroz, Raquel and Rossi, Enrico and T{\"o}rm{\"a}, P{\"a}ivi and Yang, Bohm-Jung},
  year = 2025,
  journal = {npj Quantum Mater.},
  volume = {10},
  number = {1},
  pages = {101},
  publisher = {Nature Publishing Group},
  doi = {10.1038/s41535-025-00801-3}
}

@article{PhysRevB.62.1666,
  title = {Polarization and localization in insulators: Generating function approach},
  author = {Souza, Ivo and Wilkens, Tim and Martin, Richard M.},
  journal = {Phys. Rev. B},
  volume = {62},
  issue = {3},
  pages = {1666--1683},
  numpages = {0},
  year = {2000},
  month = {Jul},
  publisher = {American Physical Society},
  doi = {10.1103/PhysRevB.62.1666},
  url = {https://link.aps.org/doi/10.1103/PhysRevB.62.1666}
}

@article{PhysRevB.56.12847,
  title = {Maximally localized generalized Wannier functions for composite energy bands},
  author = {Marzari, Nicola and Vanderbilt, David},
  journal = {Phys. Rev. B},
  volume = {56},
  issue = {20},
  pages = {12847--12865},
  numpages = {0},
  year = {1997},
  month = {Nov},
  publisher = {American Physical Society},
  doi = {10.1103/PhysRevB.56.12847},
}

@article{6gbq-d7jk,
  title = {Scaling behavior and giant field enhancement of the thermal conductivity in the honeycomb antiferromagnet ${\mathrm{BaCo}}_{2}{({\mathrm{AsO}}_{4})}_{2}$},
  author = {Hu, Jiayi and Zhong, Ruidan and Czajka, Peter and Gao, Tong and Cava, R. J. and Ong, N. P.},
  journal = {Phys. Rev. Mater.},
  volume = {9},
  issue = {6},
  pages = {L061401},
  numpages = {8},
  year = {2025},
  month = {Jun},
  publisher = {American Physical Society},
  doi = {10.1103/6gbq-d7jk},
  url = {https://link.aps.org/doi/10.1103/6gbq-d7jk}
}

@article{nanolett1c01719,
author = {Choe, Jeongheon and Lujan, David and Rodriguez-Vega, Martin and Ye, Zhipeng and Leonardo, Aritz and Quan, Jiamin and Nunley, T. Nathan and Chang, Liang-Juan and Lee, Shang-Fan and Yan, Jiaqiang and Fiete, Gregory A. and He, Rui and Li, Xiaoqin},
title = {Electron-Phonon and Spin-Lattice Coupling in Atomically Thin Layers of MnBi2Te4},
journal = {Nano Letters},
volume = {21},
number = {14},
pages = {6139-6145},
year = {2021},
doi = {10.1021/acs.nanolett.1c01719}
}

@article{PhysRevLett.116.097204,
  title = {Antiferromagnetic Spin Seebeck Effect},
  author = {Wu, Stephen M. and Zhang, Wei and KC, Amit and Borisov, Pavel and Pearson, John E. and Jiang, J. Samuel and Lederman, David and Hoffmann, Axel and Bhattacharya, Anand},
  journal = {Phys. Rev. Lett.},
  volume = {116},
  issue = {9},
  pages = {097204},
  numpages = {5},
  year = {2016},
  month = {Mar},
  publisher = {American Physical Society},
  doi = {10.1103/PhysRevLett.116.097204}
}

@article{POSTAVA1997199,
title = {Linear and quadratic magneto-optical measurements of the spin reorientation in epitaxial Fe films on MgO},
journal = {J. Magn. and Magn. Mater.},
volume = {172},
number = {3},
pages = {199-208},
year = {1997},
issn = {0304-8853},
doi = {10.1016/S0304-8853(97)00098-X},
author = {K. Postava and H. Jaffres and A. Schuhl and F. {Nguyen Van Dau} and M. Goiran and A.R. Fert},
}

@article{PhysRevB.68.064401,
  title = {Linear and quadratic magneto-optical Kerr effects in continuous and granular ultrathin monocrystalline Fe films},
  author = {Sep\'ulveda, B. and Huttel, Y. and Mart\'{\i}nez Boubeta, C. and Cebollada, A. and Armelles, G.},
  journal = {Phys. Rev. B},
  volume = {68},
  issue = {6},
  pages = {064401},
  numpages = {11},
  year = {2003},
  month = {Aug},
  publisher = {American Physical Society},
  doi = {10.1103/PhysRevB.68.064401},
  url = {https://link.aps.org/doi/10.1103/PhysRevB.68.064401}
}

@article{wohlrath2025quadratic,
  title={Quadratic magneto-optical Kerr effect spectroscopy: polarization variation method for investigation of magnetic and magneto-optical anisotropies},
  author={Wohlrath, V and Sadeghi, Z and Kim{\'a}k, J and Hovo{\v{r}}{\'a}kov{\'a}, K and Kuba{\v{s}}{\v{c}}{\'\i}k, P and Schmoranzerov{\'a}, E and N{\'a}dvorn{\'\i}k, L and Troj{\'a}nek, F and N{\v{e}}mec, P and Ostatnick{\`y}, T},
  journal={J. Phys. D: Appl. Phys.},
  volume={58},
  number={15},
  pages={155001},
  year={2025},
  doi={10.1088/1361-6463/adb760}
}

@article{nyy3-h8wr,
  title = {Ultrafast Magneto-Optical Fingerprints of Altermagnetism in MnTe},
  author = {Yang, Xu and Cheng, Xingkai and Deng, Zhuo and Gao, Yu-Han and Yang, Qing-Lin and Chang, Zheng and Yang, Peng-Tao and Feng, Hong-Mei and Zhang, Xiang-Qun and He, Wei and Liu, Junwei and Cheng, Zhao-Hua},
  journal = {Phys. Rev. Lett.},
  volume = {136},
  issue = {25},
  pages = {256702},
  numpages = {9},
  year = {2026},
  month = {Jun},
  publisher = {American Physical Society},
  doi = {10.1103/nyy3-h8wr},
}

@article{kimak2026ultrafast,
  title={Ultrafast control of spin order by linearly polarized light in noncollinear antiferromagnetic metals},
  author={Kimak, J and Nerodilova, M and Carva, K and Ghosh, S and Zelezny, J and Ostatnicky, T and Zemen, J and Johnson, F and Boldrin, D and Rendell-Bhatti, F and others},
  journal={arXiv preprint arXiv:2601.07753},
  year={2026},
  doi={10.48550/arXiv.2601.07753}
}

@article{PhysRevB.100.064403,
  title = {Quadratic magneto-optic Kerr effect spectroscopy of Fe epitaxial films on MgO(001) substrates},
  author = {Silber, Robin and Stejskal, Ond\ifmmode \check{r}\else \v{r}\fi{}ej and Beran, Luk\'a\ifmmode \check{s}\else \v{s}\fi{} and Cejpek, Petr and Anto\ifmmode \check{s}\else \v{s}\fi{}, Roman and Matalla-Wagner, Tristan and Thien, Jannis and Kuschel, Olga and Wollschl\"ager, Joachim and Veis, Martin and Kuschel, Timo and Hamrle, Jaroslav},
  journal = {Phys. Rev. B},
  volume = {100},
  issue = {6},
  pages = {064403},
  numpages = {16},
  year = {2019},
  month = {Aug},
  publisher = {American Physical Society},
  doi = {10.1103/PhysRevB.100.064403}
}

@article{berry_quantal_1984,
    author = {Berry, Michael Victor},
    title = {Quantal phase factors accompanying adiabatic changes},
    journal = {Proc. R. Soc. A},
    volume = {392},
    number = {1802},
    pages = {45-57},
    year = {1984},
    month = {03},
    issn = {0080-4630},
    doi = {10.1098/rspa.1984.0023}
}

@article{karplus1954hall,
  title={Hall effect in ferromagnetics},
  author={Karplus, Robert and Luttinger, JM},
  journal={Physical Review},
  volume={95},
  number={5},
  pages={1154},
  year={1954},
  publisher={APS},
  doi={10.1103/PhysRev.95.1154}
}

@article{xiao_berry_2010,
  title = {Berry phase effects on electronic properties},
  author = {Xiao, Di and Chang, Ming-Che and Niu, Qian},
  journal = {Rev. Mod. Phys.},
  volume = {82},
  issue = {3},
  pages = {1959--2007},
  numpages = {0},
  year = {2010},
  month = {Jul},
  publisher = {American Physical Society},
  doi = {10.1103/RevModPhys.82.1959}
}

@article{nagaosa_anomalous_2010,
  title = {Anomalous Hall effect},
  author = {Nagaosa, Naoto and Sinova, Jairo and Onoda, Shigeki and MacDonald, A. H. and Ong, N. P.},
  journal = {Rev. Mod. Phys.},
  volume = {82},
  issue = {2},
  pages = {1539--1592},
  numpages = {0},
  year = {2010},
  month = {May},
  publisher = {American Physical Society},
  doi = {10.1103/RevModPhys.82.1539}
}

@article{PhysRevLett.122.227402,
  title = {Nonreciprocal Directional Dichroism Induced by the Quantum Metric Dipole},
  author = {Gao, Yang and Xiao, Di},
  journal = {Phys. Rev. Lett.},
  volume = {122},
  issue = {22},
  pages = {227402},
  numpages = {5},
  year = {2019},
  month = {Jun},
  publisher = {American Physical Society},
  doi = {10.1103/PhysRevLett.122.227402}
}

@article{PhysRevB.111.085202,
  title = {Dielectric and optical markers originating from quantum geometry},
  author = {Chen, Wei},
  journal = {Phys. Rev. B},
  volume = {111},
  issue = {8},
  pages = {085202},
  numpages = {12},
  year = {2025},
  month = {Feb},
  publisher = {American Physical Society},
  doi = {10.1103/PhysRevB.111.085202},
  url = {https://link.aps.org/doi/10.1103/PhysRevB.111.085202}
}

@article{PhysRevB.111.205107,
  title = {Nonadiabatic quantum geometry and optical conductivity},
  author = {Resta, Raffaele},
  journal = {Phys. Rev. B},
  volume = {111},
  issue = {20},
  pages = {205107},
  numpages = {5},
  year = {2025},
  month = {May},
  publisher = {American Physical Society},
  doi = {10.1103/PhysRevB.111.205107},
  url = {https://link.aps.org/doi/10.1103/PhysRevB.111.205107}
}

@article{10.1063/1.1449436,
  title={Anisotropy of quadratic magneto-optic effects in reflection},
  author={Postava, Kamil and Hrabovsk{\`y}, David and Pi{\v{s}}tora, Jarom{\'\i}r and Fert, AR and Vi{\v{s}}{\v{n}}ovsk{\`y}, {\v{S}}tefan and Yamaguchi, T},
  journal={J. Appl. Phys.},
  volume={91},
  number={10},
  pages={7293--7295},
  year={2002},
  doi = {10.1063/1.1449436}
}

@article{PhysRevLett.87.047401,
  title = {Observation of the X-Ray Magneto-Optical Voigt Effect},
  author = {Mertins, H.-Ch. and Oppeneer, P. M. and Kune\ifmmode \check{s}\else \v{s}\fi{}, J. and Gaupp, A. and Abramsohn, D. and Sch\"afers, F.},
  journal = {Phys. Rev. Lett.},
  volume = {87},
  issue = {4},
  pages = {047401},
  numpages = {4},
  year = {2001},
  month = {Jul},
  publisher = {American Physical Society},
  doi = {10.1103/PhysRevLett.87.047401}
}

@article{peotta_superfluidity_2015,
  title = {Superfluidity in Topologically Nontrivial Flat Bands},
  author = {Peotta, Sebastiano and T{\"o}rm{\"a}, P{\"a}ivi},
  year = 2015,
  journal = {Nat. Commun.},
  volume = {6},
  number = {1},
  pages = {8944},
  publisher = {Nature Publishing Group},
  doi = {10.1038/ncomms9944}
}

\section*{Optical conductivity}\label{appoc}
The optical conductivity tensor $\boldsymbol{\sigma}$ is given by~\cite{aversa_nonlinear_1995,ahn_riemannian_2022}
\begin{equation}\label{eq:opt_cond_tensor}
\sigma_{ab}(\omega')=\frac{ie^2}{\hbar}\sum_{n\neq m}\int_{\boldsymbol{k}}\frac{f_{nm\boldsymbol{k}}\omega_{nm\boldsymbol{k}}}{\omega_{mn\boldsymbol{k}}-\omega'}\left(g_{nm\boldsymbol{k}}^{ab}-\frac{i}{2}\Omega_{nm\boldsymbol{k}}^{ab}\right).
\end{equation}
where $\int_{\boldsymbol{k}}\equiv\int d\boldsymbol{k}/(2\pi)^d$ with $d$ the system dimension, $a$ and $b$ the Cartesian coordinates, $f_{nm\boldsymbol{k}}\equiv f_{n\boldsymbol{k}}-f_{m\boldsymbol{k}}$ with $f_{n\boldsymbol{k}}$ the Fermi-Dirac distribution function, $\omega'\equiv\omega+i\eta$ with $\eta>0$ the smearing parameter, and $\hbar\omega_{nm\boldsymbol{k}}$ the energy difference between the $n$th band and $m$th band. $g_{nm\boldsymbol{k}}^{ab}$ and $\Omega_{nm\boldsymbol{k}}^{ab}$ are the matrix elements of the quantum metric tensor $\boldsymbol{g}$ and the Berry curvature tensor $\boldsymbol{\Omega}$ (i.e., the real and imaginary parts of the matrix elements of the quantum geometric tensor, respectively) and can be written as $g_{nm\boldsymbol{k}}^{ab}:=(r^a_{nm\boldsymbol{k}}r^b_{mn\boldsymbol{k}}+r^b_{nm\boldsymbol{k}}r^a_{mn\boldsymbol{k}})/2$ and $\Omega_{nm\boldsymbol{k}}^{ab}:= i(r^a_{nm\boldsymbol{k}}r^b_{mn\boldsymbol{k}}-r^b_{nm\boldsymbol{k}}r^a_{mn\boldsymbol{k}})$, where $r^a_{nm\boldsymbol{k}}$ are the matrix elements of the position operator.

While the longitudinal optical conductivities $\sigma_{xx}$ and $\sigma_{yy}$ are purely contributed by the quantum metric, the transverse optical conductivities $\sigma_{xy}$ and $\sigma_{yx}$ are contributed by both quantum metric and Berry curvature. Therefore, we divide the transverse optical conductivity $\sigma_{ab}$ into the symmetric ($\sigma_{ab}^g$) and antisymmetric ($\sigma_{ab}^\Omega$) parts, according to whether the contribution comes from the quantum metric $g$ or the Berry curvature $\Omega$. Then we have $\sigma_{xy}\equiv\sigma_{xy}^g+\sigma_{xy}^\Omega$ and $\sigma_{yx}\equiv\sigma_{xy}^g-\sigma_{xy}^\Omega$.

\section*{Quantum metric}

Many physical responses depend not only on the energy of bands but also on the wavefunctions of bands. Berry curvature describes the phase changes between wavefunctions, while the quantum metric characterizes the distance between wavefunctions, i.e., the variation of the wavefunctions themselves after excluding their phases. Quantum geometry comprehensively accounts for the changes in band wavefunctions by unifying the quantum metric and Berry curvature.

In contrast to Berry curvature, which is already well-understood, the understanding of the quantum metric is still developing. The quantum metric is typically linked to the real-space spread of the wavefunctions \cite{PhysRevB.56.12847,PhysRevB.62.1666}. Furthermore, the quantum metric is also closely related to the superfluid weight in superconductivity \cite{torma2022superconductivity}, nonlinear optical responses \cite{ahn_riemannian_2022}, etc.

In this work, since high rotational symmetry prevents the quantum metric from contributing to MOEs, the quantum metric characterizes the degree of anisotropy of the system. Thus, unconventional MOEs can be enhanced by control methods that break this high rotational symmetry, for example, by an in-plane field. On the other hand, since the quantum metric is directly related to the real-space spread of the wavefunctions and carries the dimension of length squared, one could choose moir\'e unit cells with larger areas to achieve a larger magnitude of the quantum metric~\cite{torma2022superconductivity,hu2025anomalous} and thus more remarkable unconventional MOEs.

\section*{First-principles calculations}

Our calculation of the altermagnet FeF$_3$ is based on the DFT methods using the PBE\cite{PBE} form for the GGA and HSE as implemented in the Vienna ab initio simulation package (VASP)\cite{vasp}. The Gamma scheme $k$-point mesh sizes are set to $11\times11\times5$ for FeF$_3$. The plane-wave cutoff energies are set to 500 eV for FeF$_3$. Then Wannier models are constructed using wannier90~\cite{wannier90}. In numerical calculations of the optical conductivity tensor, the $k$-mesh sizes are set to $100\times100\times100$ and the smearing parameter $\eta$ is set to 0.04 eV.

\end{document}